\documentclass[prl,twocolumn,floatfix,showpacs ]{revtex4-1}
\UseRawInputEncoding
\usepackage{amsmath}
\usepackage{amssymb}
\usepackage{graphicx}
\usepackage{dcolumn}
\usepackage{bm}
\usepackage[colorlinks=true,linkcolor=blue,anchorcolor=blue, citecolor=cyan,urlcolor=cyan]{hyperref}
\usepackage[mathlines]{lineno}
\usepackage{ulem}
\usepackage{pifont}
\usepackage{epstopdf}
\usepackage{tabularx} 
\begin{document}
\title{External-field-induced altermagnetism in  experimentally synthesized monolayer  $\mathrm{CrX_3}$ (X=Cl, Br and I)}
\author{San-Dong Guo}
\email{sandongyuwang@163.com}
\affiliation{School of Electronic Engineering, Xi'an University of Posts and Telecommunications, Xi'an 710121, China}

\begin{abstract}
Net-zero-magnetization magnets are attracting significant research interest, driven by their potential for ultrahigh density and ultrafast performance.
Among these materials, the altermagnets possess alternating spin-splitting band structures and exhibit a range of phenomena previously thought to be exclusive to ferromagnets,  including  the anomalous Hall and Nernst effects, non-relativistic spin-polarized currents, and the magneto-optical Kerr effect.
Bulk altermagnets have been experimentally identified, while two-dimensional (2D) altermagnets remain experimentally unexplored.
Here, we take experimentally synthesized 2D ferromagnetic  $\mathrm{CrX_3}$ (X=Cl, Br and I) as the parent material and achieve altermagnetism through external field. First, we achieve the transition from ferromagnetism to antiferromagnetism through biaxial strain. Subsequently, we break the space inversion symmetry  while preserving the mirror symmetry via an electric field, thereby inducing altermagnetism.
Moreover, through the application of Janus engineering to construct  $\mathrm{CrX_{1.5}Y_{1.5}}$ (X$\neq$Y=Cl, Br and I), the phase transition from ferromagnetism to antiferromagnetism induced by strain alone is sufficient to trigger the emergence of altermagnetism. All six monolayers possess the symmetry of $i$-wave spin-splitting.
The computational results suggest that   $\mathrm{CrCl_3}$ can be readily tuned to exhibit altermagnetism through external field  in experiment, thanks to its low strain threshold for magnetic phase transition. Our work provides experimentally viable materials and methods for realizing altermagnetism, which can advance the development of 2D altermagnetism.

\end{abstract}
\maketitle
\textcolor[rgb]{0.00,0.00,1.00}{\textbf{Introduction.---}}
Although ferromagnets play a significant role in spintronics, they face limitations such as stray field problems and power consumption issues\cite{k1,k1-1,k2}. In contrast, contemporary research in spintronics has increasingly focused on net-zero-magnetization magnets. These magnets offer several advantages over ferromagnets for spintronic devices, including higher storage density, enhanced robustness against external magnetic fields, and faster writing speeds, all of which stem from their net-zero magnetic moment \cite{k1,k1-1,k2}. Typical examples of net-zero-magnetization magnets include $PT$-antiferromagnets (which possess combined space inversion ($P$) and time-reversal ($T$) symmetry ($PT$)), altermagnets  and fully-compensated ferrimagnets\cite{k4,k5}.  Unlike $PT$-antiferromagnet,  both altermagnet and fully-compensated ferrimagnet  can  produce anomalous Halll/Nernst effect  and magneto-optical Kerr effect\cite{k4,h13,f4}.

Recently, altermagnetism has become a highly focused hot topic in the fields of condensed matter physics and materials science\cite{h13}.
The  altermagnetism is distinguished by several key features: it exhibits robust breaking of $T$ symmetry, displays antiparallel magnetic order,  possesses alternating spin-split band structures and has vanishing net magnetization due to symmetry constraints.
The momentum-dependent spin-splitting of $d$-, $g$-, or $i$-wave symmetry  in Brillouin zone (BZ) does not require the assistance  of spin-orbital coupling (SOC)\cite{k4,k5,k511,k512,k513}. Recently, twisted altermagnetism has been proposed in twisted magnetic van der Waals (vdW) bilayers, which utilize one of the five  two-dimensional (2D) Bravais lattices\cite{k8,k80}. In these systems, an out-of-plane electric field can induce valley polarization due to valley-layer coupling \cite{k9,k10,k910}. Moreover, an antiferroelectric altermagnet has been proposed, where antiferroelectricity and altermagnetism coexist in a single material\cite{k7-3-2}.
Experimentally, several bulk  altermagnetic materials have been identified that exhibit momentum-dependent spin-splitting\cite{k4,h13}.
Nevertheless, 2D altermagnets have been reported only in theory \cite{h13,k60,k6,k7,k7-1,k7-2,k7-3} and remain experimentally unexplored.
Therefore, it is important and meaningful to provide experimentally feasible materials and strategies to realize  altermagnetism.

In addition to intrinsic altermagnet, it can also be achieved through external-field-tuned $PT$-antiferromagnets. The $PT$-antiferromagnet exhibits spin degeneracy, and many interesting physical phenomena and effects are prohibited.
The spin-splitting can be achieved by arranging magnetic atoms with opposite spin polarizations in distinct environments\cite{gsd}. Essentially, this requires breaking the system's $P$ symmetry\cite{qq1,qq2,qq3}.  If $PT$-antiferromagnet only possesses $P$ symmetry, it will transform into fully-compensated ferrimagnet upon $P$ symmetry breaking\cite{qq3}.  If $PT$-antiferromagnet not only possesses $P$ symmetry but also has rotational or mirror ($C$ or $M$) symmetry, it will become altermagnet after the breaking of $P$ symmetry\cite{qq1,qq2,qq3}.  Here, $P$, $C$ and $M$ require the connection of magnetic atoms with opposite spin polarizations.   Here, we will achieve the altermagnet by inducing $\mathrm{CrX_3}$ (X=Cl, Br and I) to transition from ferromagnetic (FM) to antiferromagnetic (AFM) state by strain,  and then  breaking their  $P$ symmetry through electric field.

 \begin{figure}[t]
    \centering
    \includegraphics[width=0.48\textwidth]{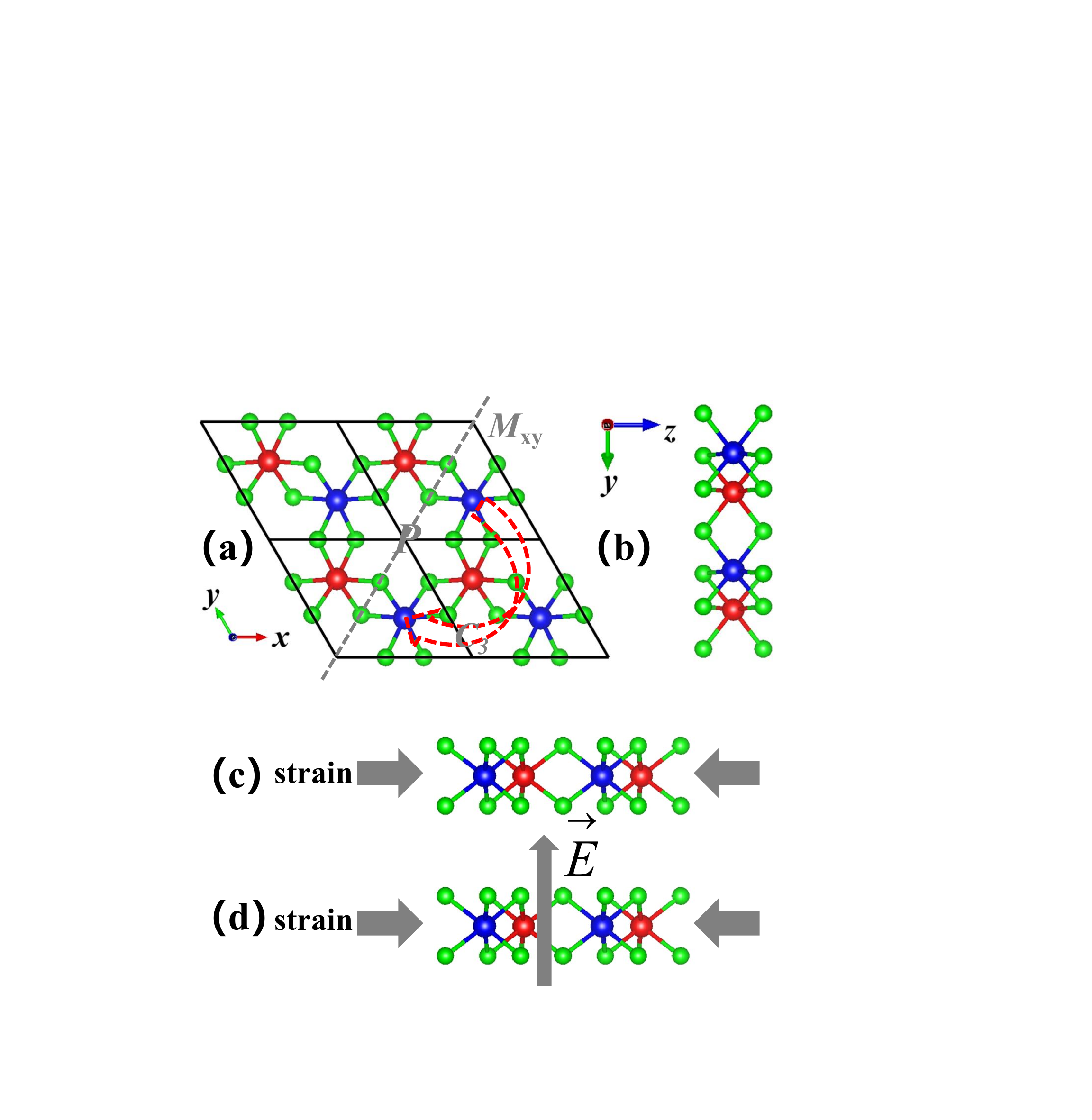}
    \caption{(Color online) (a and b): the crystal structures of $\mathrm{CrX_3}$,  and the blue/red and green balls  represent Cr  and
      X atoms,  respectively. The small black rhombus represents the primitive unit cell; (c): when the appropriate strain is applied, a phase transition from ferromagnetism to $PT$-antiferromagnetism occurs; (d): when an additional electric field is applied, a phase transition from $PT$-antiferromagnetism to altermagnetsim is induced.}\label{a}
\end{figure}
\begin{table}
    \centering
    \caption{For  $\mathrm{CrX_3}$  and $\mathrm{CrX_{1.5}Y_{1.5}}$, the  lattice constants $a$ ($\mathrm{{\AA}}$), the energy band gap $Gap$ (eV),  the critical value of $a/a_0$ ($(a-a_0)/a_0$) corresponding to this phase transition from FM to AFM1 ordeing, and the Young's modulus $C_{2D}$ ($\mathrm{Nm^{-1}}$).}
    \label{tab}
    \begin{tabularx}{0.45\textwidth}{@{\extracolsep{\fill}}ccccc} 
        \hline\hline
        Monolayer & $a$&$Gap$&$a/a_0$ &$C_{2D}$\\ \hline
         $\mathrm{CrCl_3}$ & 6.06&1.767&0.977 (-2.3\%)& 34\\ \hline
         $\mathrm{CrBr_3}$  & 6.44&1.541 &0.962 (-3.8\%)& 29\\ \hline
         $\mathrm{CrI_3}$ & 7.00&1.296 &0.944 (-5.6\%)& 23\\ \hline
       $\mathrm{CrCl_{1.5}Br_{1.5}}$  & 6.26 &1.595&0.967 (-3.3\%)&33\\ \hline
$\mathrm{CrCl_{1.5}I_{1.5}}$  & 6.59& 1.255 &0.951 (-4.9\%)&32\\ \hline
       $\mathrm{CrBr_{1.5}I_{1.5}}$  & 6.74 &1.335&0.949 (-5.1\%)&28\\ \hline\hline
    \end{tabularx}
\end{table}
\begin{figure}[t]
    \centering
    \includegraphics[width=0.40\textwidth]{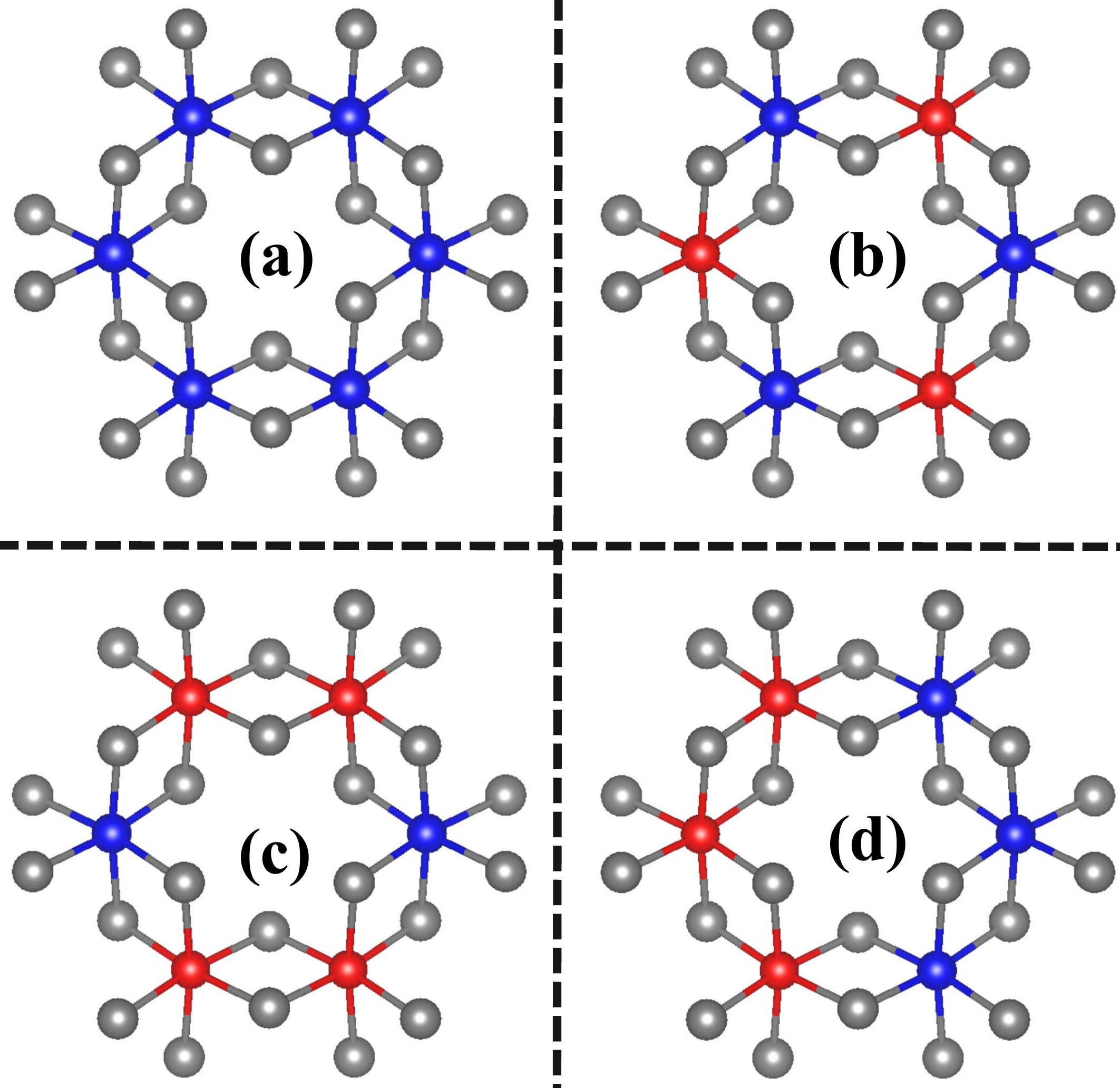}
    \caption{(Color online) Four magnetic configurations: FM (a),  AFM1 (b), AFM2 (c) and AFM3 (d), and the blue and red balls represent spin-up and spin-down atoms, respectively. }\label{b}
\end{figure}

\begin{figure*}[t]
    \centering
    \includegraphics[width=0.9\textwidth]{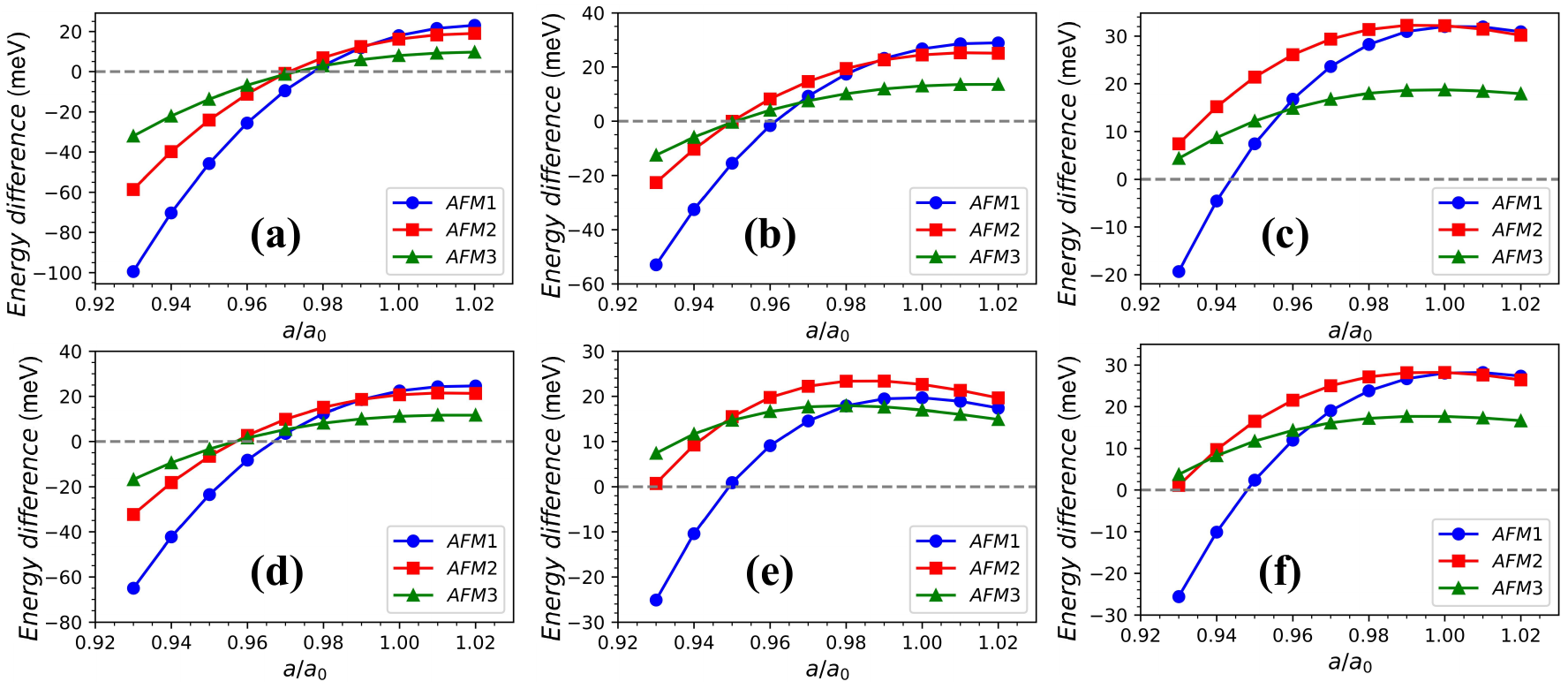}
     \caption{(Color online) Calculated energy differences of AFM1, AFM2 and AFM3  with respect to  FM state as a function of strain ($a/a_0$) for $\mathrm{CrCl_3}$ (a), $\mathrm{CrBr_3}$ (b), $\mathrm{CrI_3}$ (c), $\mathrm{CrCl_{1.5}Br_{1.5}}$ (d), $\mathrm{CrCl_{1.5}I_{1.5}}$ (e) and $\mathrm{CrBr_{1.5}I_{1.5}}$ (f). }\label{c}
\end{figure*}
\textcolor[rgb]{0.00,0.00,1.00}{\textbf{Crystal structures and Approach.---}}
Monolayer $\mathrm{CrX_3}$ (X=Cl, Br and I) have already been synthesized experimentally\cite{p1,p2,p3}.
\autoref{a}(a, b) illustrate the crystal structures of monolayer $\mathrm{CrX_3}$, which is composed of three monatomic planes in the sequence X-Cr-X. The unit cell contains  two Cr atoms, which provides the structural condition for the realization  of altermagnetism.
 The $\mathrm{CrX_3}$ crystallizes in the  $P\bar{3}1m$ space group (No.162).
 It is evident that certain symmetry operations can map one Cr sublattice onto the other (see \autoref{a} (a)). These operations include inversion center $P$, which is situated between the nearest-neighbor Cr-Cr bonds and at the centers of the Cr hexagons. Additionally, vertical mirrors  $M$ (for example $M_{xy}$) perpendicular to the Cr-Cr bonds and passing through the inversion center also play a role. However, three-fold proper rotations  $C_3$ connect atoms within the same Cr sublattice.

 Two Cr sublattices can be connected through $M$ symmetry, which provides the symmetrical condition for realizing  altermagnetism. However, the presence of the $P$ symmetry   limits the existence of altermagnetism. Therefore, in order to realize altermagnetism, it is necessary to retain $M$ symmetry while breaking the $P$ symmetry.  The electric field is a highly effective mean of breaking $P$ symmetry in 2D materials\cite{zg4},  which is a volatile regulation.
 It should be noted  that if the two magnetic sublattices are not connected by $M$ or $C$ symmetry, then after the electric field breaks the $P$ symmetry, the material may become a fully compensated ferrimagnet\cite{f4,k910}.

  In fact, the externally applied electric field can be equivalently replaced by the built-in electric field induced by the Janus structure. That is to say, in addition to electric field, Janus engineering is also an effective method for breaking $P$ symmetry in 2D sandwich structures.  Janus monolayer   $\mathrm{CrX_{1.5}Y_{1.5}}$ (X$\neq$Y=Cl, Br and I) can be constructed  by  replacing one of two X  layers with Y  atoms in monolayer  $\mathrm{CrX_3}$,  giving rise to  three Janus monolayers  $\mathrm{CrCl_{1.5}Br_{1.5}}$, $\mathrm{CrCl_{1.5}I_{1.5}}$ and $\mathrm{CrBr_{1.5}I_{1.5}}$.
Monolayer  $\mathrm{CrX_{1.5}Y_{1.5}}$ loses $P$ and horizontal $M$ symmetries with   $P31m$ space group  (No.157), but retains the vertical $M$ symmetry.

​In addition to the constraints of structural symmetry, $\mathrm{CrX_3}$ should exhibit a N$\acute{e}$el  AFM ground state to achieve altermagnetism.
 Unfortunately, it has been demonstrated that $\mathrm{CrX_3}$ possesses ferromagnetism  both theoretically and experimentally\cite{p1,p2,p3,p4,p5,p6}. However, strain is a very effective way to tune the magnetic properties of 2D materials.
 For  $\mathrm{CrX_3}$, strain combined with an electric field should be employed to induce altermagnetism. Initially, we will apply biaxial strain to induce a phase transition from ferromagnetism to antiferromagnetism  \autoref{a} (c). Subsequently, an electric field will be applied to achieve a altermagnetic state  \autoref{a} (d). For Janus   $\mathrm{CrX_{1.5}Y_{1.5}}$,  due to the broken $P$ symmetry, strain alone is sufficient to achieve  altermagnetism from ferromagnetism.

\textcolor[rgb]{0.00,0.00,1.00}{\textbf{Computational detail.---}}
Within density functional theory (DFT) \cite{1}, we perform the  spin-polarized  first-principles calculations  by using the projector augmented-wave (PAW) method,  as implemented in Vienna ab initio simulation package (VASP)\cite{pv1,pv2,pv3}.  The  generalized gradient approximation (GGA) of  Perdew, Burke, and Ernzerhof (PBE)\cite{pbe} is adopted as the exchange-correlation functional to investigate $\mathrm{CrX_3}$ and  $\mathrm{CrX_{1.5}Y_{1.5}}$\cite{p4}. The kinetic energy cutoff  of 500 eV,  total energy  convergence criterion of  $10^{-8}$ eV, and  force convergence criterion of  0.0001 $\mathrm{eV.{\AA}^{-1}}$ are adopted  to obtain the reliable results.
 To avoid interlayer interactions,  a slab model is used  with a vacuum thickness of more than 15 $\mathrm{{\AA}}$ along $z$ direction.
The  BZ is sampled  with a 12$\times$12$\times$1 Monkhorst-Pack $k$-point meshes  for structure relaxation and electronic structure calculations.
 To calculate Young's modulus,  we calculate the elastic stiffness tensor  $C_{ij}$   by using strain-stress relationship (SSR) method, which  have been renormalized by   $C^{2D}_{ij}$=$L_z$$C^{3D}_{ij}$ with  the $L_z$ being the length of unit cell along $z$ direction.

\begin{figure*}[t]
    \centering
    \includegraphics[width=0.9\textwidth]{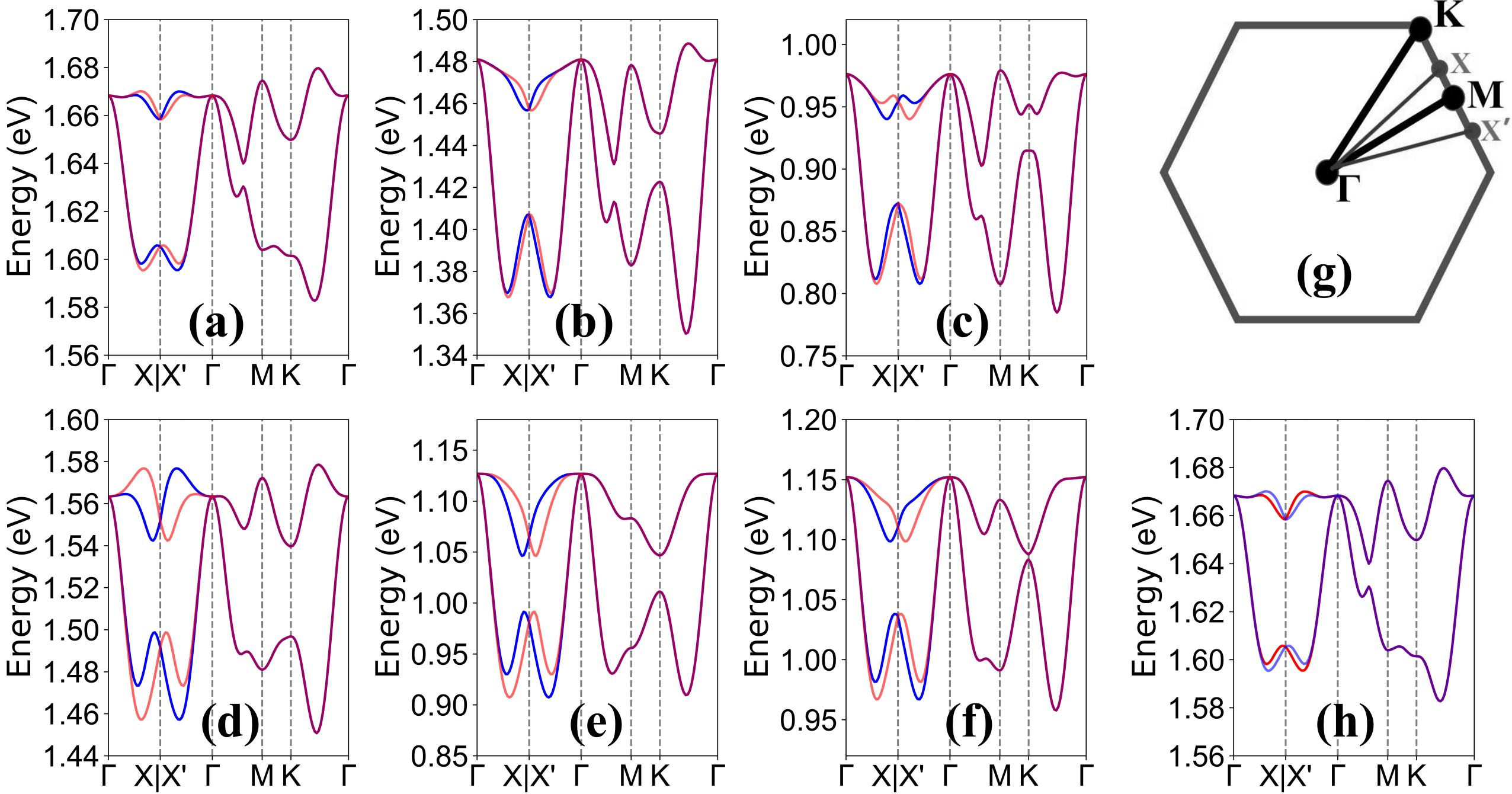}
     \caption{(Color online) The spin-polarized  energy band structures of   strained  $\mathrm{CrCl_3}$ (a), $\mathrm{CrBr_3}$ (b), $\mathrm{CrI_3}$ (c), $\mathrm{CrCl_{1.5}Br_{1.5}}$ (d), $\mathrm{CrCl_{1.5}I_{1.5}}$ (e) and $\mathrm{CrBr_{1.5}I_{1.5}}$ (f) in the conduction bands near the Fermi energy level, and  an electric field of $E$=+0.20$\mathrm{V/{\AA}}$  is applied for (a), (b) and (c).   The (g) shows   the BZ   with high symmetry points, and the (h) shows spin-polarized  energy band structures of    $\mathrm{CrCl_3}$ with $E$=-0.20$\mathrm{V/{\AA}}$.  In (a, b, c, d, e, f, h), the $a/a_0$ is 0.96, 0.95, 0.94, 0.95, 0.94, 0.94 and 0.96, respectively. The spin-up and spin-down channels are depicted in blue and red, and the  purple color means spin degeneracy. }\label{d}
\end{figure*}

\begin{figure*}[t]
    \centering
    \includegraphics[width=0.9\textwidth]{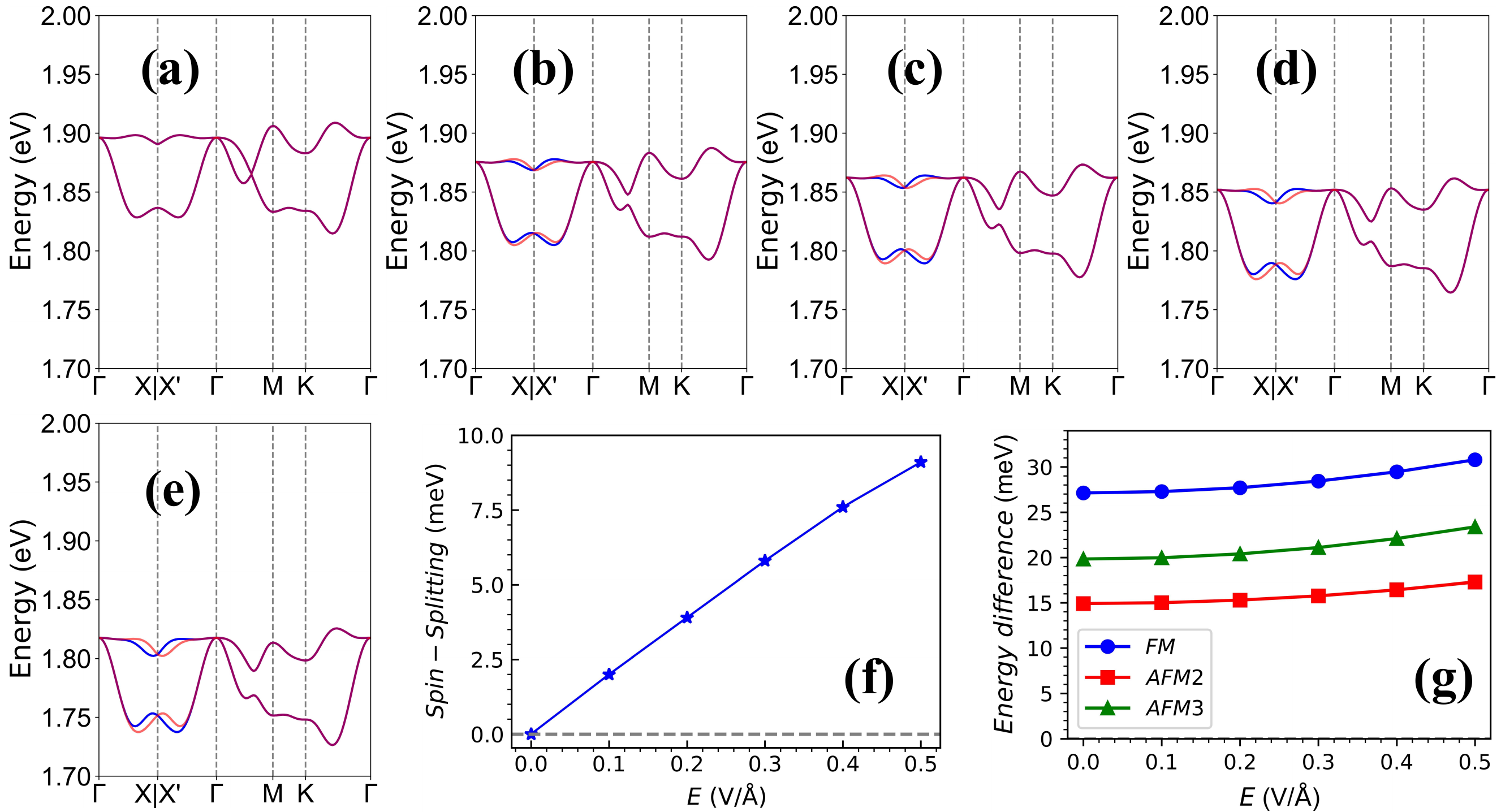}
     \caption{(Color online) For   $\mathrm{CrCl_3}$  with $a/a_0$ being 0.96, the spin-polarized  energy band structures  at $E$=+0.00 (a), +0.20 (b), +0.30 (c), +0.40 (d) and +0.50 (e) $\mathrm{V/{\AA}}$;  The (f) shows  the spin-splitting of the first two conduction bands at one point along the $\Gamma$-X path, and the (g) shows the energy differences of FM, AFM2 and AFM3  with respect to  AFM1 state as a function of $E$.  In (a, b, c, d, e), the spin-up and spin-down channels are depicted in blue and red, and the  purple color means spin degeneracy. }\label{e}
\end{figure*}
\begin{figure}[t]
    \centering
    \includegraphics[width=0.45\textwidth]{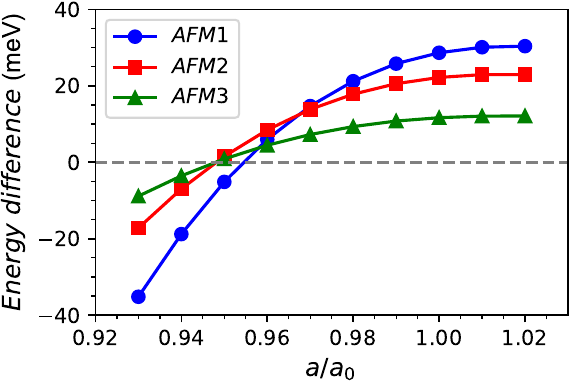}
     \caption{(Color online) For   $\mathrm{CrCl_3}$, the energy differences of AFM1, AFM2 and AFM3  with respect to  FM state as a function of strain ($a/a_0$) with GGA+$U$ ($U$=2.63 eV\cite{p5}). }\label{f}
\end{figure}

\textcolor[rgb]{0.00,0.00,1.00}{\textbf{Main results.---}}
The possible four  magnetic configurations (see \autoref{b}) are considered to  evaluate the magnetic ground state of monolayer   $\mathrm{CrX_3}$ and  $\mathrm{CrX_{1.5}Y_{1.5}}$,  including FM   and  three  AFM  (AFM1, AFM2 and AFM3) configurations. The  AFM1 ordering corresponds to the so-called N$\acute{e}$el  AFM state.   In the original works, only the N$\acute{e}$el  AFM  configuration is  considered\cite{p4,p5,p6}. Here, we consider three AFM configurations in order to accurately determine the magnetic ground state during the process of applying strain.
Calculated results indicate that all six monolayers are FM ground states, with the corresponding optimized lattice constants listed in \autoref{tab}, which agree well with available  results\cite{p4,p5,p6}.    The energy band structures of  six monolayers are calculated, which are plotted in
FIG.S1\cite{bc}, and the corresponding energy band gaps are summarized in \autoref{tab}.

Here, the ratio $a/a_0$ (ranging from 0.93 to 1.02) is employed to simulate biaxial strain, where $a$ and $a_0$
denote the strained and unstrained lattice constants, respectively. Specifically, $a/a_0$$<$1 signifies compressive strain, while $a/a_0$$>$1 indicates tensile strain. To begin with, we ascertain the magnetic ground state of strained $\mathrm{CrX_3}$ and  $\mathrm{CrX_{1.5}Y_{1.5}}$
​by comparing the energy differences between the  AFM1/AFM2/AFM3 and FM configurations, as illustrated in \autoref{c}.
As $a/a_0$ decreases (increasing the compressive
strain), the energy diﬀerence between AFM1/AFM2/AFM3 and FM configurations  decreases.  It is evident that the energy difference between the AFM1 and FM orders decreases more rapidly. A phase transition to the AFM1 state occurs when the energy difference between AFM1 and FM ordering becomes less than zero.
The critical values of $a/a_0$ ($(a-a_0)/a_0$) corresponding to this phase transition are summarized in \autoref{tab}.

 For $\mathrm{CrX_3}$,   the critical values of the magnetic phase transition are consistent with the original computational results ( -2.5\%, -4.1 \% and -5.7 \%  for the  $\mathrm{CrCl_3}$, $\mathrm{CrBr_3}$ and $\mathrm{CrI_3}$\cite{p4}).
It is found that the heavier the atomic mass contained in a monolayer, the greater the compressive strain required (the smaller the $a/a_0$).
That is to say,  $\mathrm{CrCl_3}$ is more likely to become AFM1 ordering with $a/a_0$  ($(a-a_0)/a_0$) for 0.977 (-2.3\%).
Therefore, strain can indeed achieve the magnetic configuration that satisfy the requirements of  altermagnetism.

The spin-polarized  energy band structures of   strained  $\mathrm{CrX_3}$ are plotted in FIG.S2\cite{bc},  and they are all semiconductors with spin degeneracy caused by $PT$ symmetry. We now discuss possible routes to achieve the symmetry breaking by an  out-of-plane electric field. Such electric filed breaks the $PT$ symmetry by the introduction  of an electrostatic potential, making  the two X layers crystallographically inequivalent.  The crystal symmetry changes  from $P\bar{3}1m$  (No.162) to  $P31m$  (No.157).  The spin-polarized  energy band structures of   strained  $\mathrm{CrX_3}$ with   an electric field of $E$=+0.20$\mathrm{V/{\AA}}$ are plotted in FIG.S3\cite{bc} and \autoref{d} (a, b, c). Compare with the band structures without an electric field, it is clearly seen that there is spin-splitting along the $\Gamma$-X and X'-$\Gamma$ paths, which are connected through   [$C_2$$\parallel$$M$] symmetry, giving rise to  altermagnetism.
When the direction of the electric field is reversed, the order of spin-splitting is correspondingly inverted (see \autoref{d} (h)).

Below, we discuss the symmetry of spin-splitting in  altermagnetic $\mathrm{CrX_3}$, achieved through strain and electric field.
The $PT$ symmetry of  $\mathrm{CrX_3}$ is broken if the two X layers are not equivalent caused by electric field, which leads to that the band structures in momentum space only exhibit $C_3$ symmetry. However,  the electronic structure in momentum space for a given spin must be inversion symmetric without including SOC, which make the system possess $C_6$ symmetry. Moreover, in a 2D system, the electronic band structures depend solely on $k_x$ and $k_y$, and are independent of $k_z$. Therefore, the  $i$-wave symmetry of spin-splitting can be realized by strain and electric field in  monolayer $\mathrm{CrX_3}$.
This implies that there are six nodal lines of spin-splitting in the first BZ. Similar results can be found in  $\mathrm{MnPSe_3}$ and $\mathrm{CrCS_3}$\cite{qq1,qq3}.

The two X layers can be also made inequivalent by Janus  engineering, giving rise to $\mathrm{CrX_{1.5}Y_{1.5}}$ with built-in electric field. Such Janus monolayer  MoSSe has already been synthesized experimentally\cite{mos}, which provides a reference for the synthesis of  $\mathrm{CrX_{1.5}Y_{1.5}}$.
The spin-polarized  energy band structures of   strained  $\mathrm{CrX_{1.5}Y_{1.5}}$  are plotted in FIG.S4\cite{bc} and \autoref{d} (d, e, f).  Without applied electric field,  there is spin-splitting along the $\Gamma$-X and X'-$\Gamma$ paths,  producing  altermagnetism.
Like $\mathrm{CrX_3}$ under an applied electric field, strained  $\mathrm{CrX_{1.5}Y_{1.5}}$ also exhibits $i$-wave spin-splitting symmetry.
The spin-splitting in the Janus structure is more pronounced than that in the parent monolayer under an applied electric field, which may be attributed to the large built-in electric field (see FIG.S5\cite{bc}).

Finally, we take $\mathrm{CrCl_3}$ as an example to consider the effect of the electric field magnitude on spin-splitting.
The spin-polarized  energy band structures  of  $\mathrm{CrCl_3}$ with $a/a_0$ being 0.96 at $E$=+0.00, +0.20, +0.30, +0.40 and +0.50 $\mathrm{V/{\AA}}$ are plotted in \autoref{e} (a, b, c, d, e), and   the spin-splitting of the first two conduction bands at one point along the $\Gamma$-X path are show in \autoref{e} (f).
It is found that the spin-splitting gradually increases with the increase of electric field.  At $E$=0.40 $\mathrm{V/{\AA}}$, the spin-splitting reaches 7.6 meV.
Recently, an intense electric field exceeding 0.4 $\mathrm{V/{\AA}}$ has been achieved in 2D materials through dual ionic gating\cite{zg7}, which offers the potential to realize large altermagnetic spin-splitting in  the $\mathrm{CrX_3}$.
Another important factor for achieving altermagnetism  is that  $\mathrm{CrCl_3}$ maintains the AFM1 ordering under an applied electric field.
The energy differences between the FM/AFM2/AFM3 and AFM1 configurations are plotted in \autoref{e} (g). Within the considered range of electric field, all the energy differences are positive, implying that $\mathrm{CrCl_3}$ remains in the AFM1 state.

\textcolor[rgb]{0.00,0.00,1.00}{\textbf{Discussion and conclusion.---}}
 To elucidate
mechanical performance of  $\mathrm{CrX_3}$ and  $\mathrm{CrX_{1.5}Y_{1.5}}$, the in-plane Young's modulus $C_{2D}$ are calculated based on  $C_{ij}$\cite{ela}, which are listed in \autoref{tab}. The obtained $C_{2D}$ of six monolayers  are much smaller than graphene ($\sim 340\pm 40$ Nm$^{-1}$) and MoS$_2$ ($\sim 126.2$ Nm$^{-1}$\cite{q5-1,q5-1-1}), indicating their better mechanical flexibility. This is in favour of experimentally realizing the phase transition from FM to， AFM1 phase  by strain, further producing  altermagnetism. The  strain as large as 5.6\% for $\mathrm{MoS_2}$  has been realized experimentally\cite{gsd3}. So, the  0.977 (-2.3\%) strain can be easily achieved  experimentally due to small  Young's modulus in $\mathrm{CrCl_3}$.

The reliability of our results essentially depends on whether the strain will induce a transition from ferromagnetism to the AFM1 state.
A previous theoretical study of bulk $\mathrm{CrX_3}$ (X=Cl, Br and I)
 crystals shows  that FM ordering can be obtained in agreement with experimental results for $\mathrm{CrBr_3}$  and $\mathrm{CrI_3}$  using the GGA method\cite{jpcm}. However, the AFM state of $\mathrm{CrCl_3}$  can only be accurately reproduced by incorporating on-site electron-electron repulsion. Therefore, we use monolayer $\mathrm{CrCl_3}$ as an example to illustrate the impact of electron correlation effects ($U$=2.63 eV\cite{p5}) on magnetic phase transitions.  The energy difference between GGA and GGA+$U$ (see \autoref{c} (a) and \autoref{f}) shows a similar trend with the variation of strain. With GGA+$U$, strain can still achieve the transition from ferromagnetism to the AFM1 state, and the critical value of $a/a_0$  ($(a-a_0)/a_0$) corresponding to magnetic  phase transition becomes 0.956 (-4.4\%). Thus, our proposed method can indeed achieve altermagnetism  in  monolayer
Cr-trihalide and its Janus structures.

In summary, we use 2D FM $\mathrm{CrX_3}$ (X=Cl, Br and I) to achieve altermagnetism via external fields. Biaxial strain induces a transition from ferromagnetism to antiferromagnetism.  And then, an electric field breaks $P$ symmetry while preserving $M$ symmetry, inducing altermagnetism. Janus engineering also enables altermagnetism through strain-induced  magnetic phase transitions. The $\mathrm{CrCl_3}$ is particularly promising due to its low strain threshold for magnetic phase transition. Our works provide feasible materials and methods for realizing 2D altermagnetism.

\begin{acknowledgments}
This work is supported by Natural Science Basis Research Plan in Shaanxi Province of China  (2021JM-456). We are grateful to Shanxi Supercomputing Center of China, and the calculations were performed on TianHe-2. We thank Prof. Guangzhao Wang for providing VASP software and helpful discussions.
\end{acknowledgments}


\begin{references}

\bibitem{k1}X. Hu, Half-metallic antiferromagnet as a prospective material for spintronics, Adv. Mater. \textbf{24}, 294 (2012).

\bibitem{k1-1}T. Jungwirth, X. Marti,  P. Wadley and J.  Wunderlich,  Antiferromagnetic spintronics,  Nat. Nanotechnol. \textbf{11}, 231 (2016).

\bibitem{k2}T. Jungwirth, J. Sinova, A. Manchon, X. Marti, J. Wunderlich
and C. Felser, The multiple directions of antiferromagnetic spintronics, Nat. Phys. \textbf{14}, 200 (2018).


\bibitem{k4}L. $\mathrm{\breve{S}}$mejkal, J. Sinova and T. Jungwirth, Beyond conventional ferromagnetism and antiferromagnetism: A phase with nonrelativistic spin and crystal rotation symmetry,  Phys. Rev. X
\textbf{12}, 031042 (2022).


\bibitem{k5}I. Mazin, Altermagnetism-a new punch line of fundamental magnetism,  Phys. Rev. X \textbf{12}, 040002 (2022).

\bibitem{h13}L. Bai, W. Feng, S. Liu, L. $\mathrm{\breve{S}}$mejkal, Y. Mokrousov, and Y. Yao, Altermagnetism: Exploring New Frontiers in Magnetism and Spintronics, Adv. Funct. Mater.  \textbf{34}, 2409327 (2024).


\bibitem{f4}Y. Liu, S. D. Guo, Y. Li and C. C. Liu, Two-dimensional fully-compensated Ferrimagnetism, Phys. Rev. Lett. \textbf{134}, 116703 (2025).



\bibitem{k511}S. Hayami, Y. Yanagi and H. Kusunose, Momentum-Dependent Spin Splitting by Collinear Antiferromagnetic Ordering, J. Phys. Soc. Jpn. \textbf{88}, 123702 (2019).


\bibitem{k512}S. Hayami, Y. Yanagi and H. Kusunose, Bottom-up design of spin-split and reshaped electronic band structures in antiferromagnets without spin-orbit coupling: Procedure on the basis of augmented multipoles, Phys. Rev. B \textbf{102}, 144441  (2020).


\bibitem{k513}S. Hayami and H. Kusunose, Essential role of the anisotropic magnetic dipole in the anomalous Hall effect, Phys. Rev. B \textbf{103}, L180407 (2021).


\bibitem{k8}Y. Liu, J. Yu and C. C. Liu, Twisted Magnetic Van der Waals Bilayers: An Ideal Platform for Altermagnetism, Phys. Rev. Lett. \textbf{133}, 206702 (2024).

\bibitem{k80}R. He, D. Wang, N. Luo, J. Zeng,  K. Q. Chen  and L. M. Tang, Nonrelativistic Spin-Momentum Coupling in Antiferromagnetic Twisted Bilayers, Phys. Rev. Lett. \textbf{130}, 046401 (2023).


\bibitem{k9}S. D. Guo, Y. Liu, J. Yu and C. C. Liu, Valley  polarization  in twisted altermagnetism,  Phys. Rev. B \textbf{110}, L220402 (2024).


\bibitem{k10}R. W. Zhang, C. X. Cui,  R. Z. Li, J. Y. Duan, L. Li, Z. M. Yu and Y. G. Yao, Predictable gate-field control of spin in altermagnets with spin-layer coupling, Phys. Rev.
Lett.  \textbf{133}, 056401 (2024).

\bibitem{k910}S. D. Guo, Valley polarization in two-dimensional zero-net-magnetization magnets, Appl. Phys. Lett. \textbf{126}, 080502 (2025).

\bibitem{k7-3-2}X. Duan, J. Zhang, Z. Zhang, I $\breve{Z}$uti$\acute{c}$ and T. Zhou, Antiferroelectric Altermagnets: Antiferroelectricity Alters Magnets, Phys. Rev. Lett. \textbf{134}, 106801 (2025).


\bibitem{k60}S. D. Guo, X. S. Guo and G. Wang, Valley polarization in two-dimensional tetragonal altermagnetism, Phys. Rev. B  \textbf{110}, 184408 (2024).



\bibitem{k6}H.-Y. Ma, M. L. Hu, N. N. Li, J. P. Liu, W.
Yao, J. F. Jia and J. W.  Liu, Multifunctional antiferromagnetic materials with giant piezomagnetism and noncollinear spin current, Nat. Commun. \textbf{12}, 2846 (2021).

\bibitem{k7}S.-D. Guo, X.-S. Guo, K. Cheng, K. Wang, and Y. S.
Ang, Piezoelectric altermagnetism and spin-valley polarization in Janus monolayer $\mathrm{Cr_2SO}$,  Appl. Phys. Lett \textbf{123}, 082401 (2023).

\bibitem{k7-1}X. Chen, D. Wang, L. Y. Li and B. Sanyal, Giant spin-splitting and tunable spin-momentum locked transport in room temperature collinear antiferromagnetic semimetallic CrO monolayer, Appl. Phys. Lett. \textbf{123}, 022402 (2023).

\bibitem{k7-2}Y. Zhu, T.  Chen, Y. Li, L. Qiao, X. Ma, C. Liu, T. Hu, H. Gao and W. Ren, Multipiezo Effect in Altermagnetic $\mathrm{V_2SeTeO}$ Monolayer, Nano Lett. \textbf{24}, 472 (2024).

\bibitem{k7-3}Y. Wu, L. Deng, X. Yin, J. Tong, F. Tian and X. Zhang, Valley-Related Multipiezo Effect and Noncollinear Spin Current in an Altermagnet $\mathrm{Fe_2Se_2O}$ Monolayer, Nano Lett. \textbf{24}, 10534 (2024).

\bibitem{gsd}S. D. Guo, Y. L. Tao, G. Wang and Y. S. Ang, How to produce spin-splitting in antiferromagnetic materials, J. Phys.: Condens. Matter 36, 215804 (2024).

\bibitem{qq1}I. Mazin,  R. Gonz$\mathrm{\acute{a}}$lez-Hern$\mathrm{\acute{a}}$ndez  and L. $\mathrm{\breve{S}}$mejkal, Induced Monolayer Altermagnetism in  $\mathrm{MnP(S,Se)_3}$ and FeSe, arXiv:2309.02355 (2023).

\bibitem{qq2}P. Zhou, X. N. Peng, Y. Z. Hu,  B. R. Pan, S. M. Liu, Pengbo Lyu
and L. Z. Sun, Transition from Antiferromagnetism to Altermagnetism: Symmetry-Breaking Theory, arXiv:2410.17747 (2024).

\bibitem{qq3}S. D. Guo, X. S. Guo and G. Wang, Symmetry-breaking induced transition among net-zero-magnetization magnets,  arXiv:2501.06829 (2025).


\bibitem{p1}H. H. Kim, B. Yang, S. Li  et al.,  Evolution of interlayer and intralayer magnetism in
three atomically thin chromium trihalides, Proc. Natl. Acad. Sci.
U. S. A. \textbf{116}, 11131 (2019).

\bibitem{p2}B. Huang, G.  Clark,  E.  Navarro-Moratalla  et al.,  Layer-dependent ferromagnetism in a van der Waals crystal down to the monolayer limit,  Nature \textbf{546}, 270  (2017).


\bibitem{p3}S. Lu,  D.  Guo,  Z. Cheng  et al., Controllable dimensionality conversion between 1D and 2D $\mathrm{CrCl_3}$ magnetic nanostructures, Nat. Commun. \textbf{14}, 2465 (2023).

\bibitem{zg4}H. J. Zhao,  X. Liu,  Y. Wang, Y. Yang, L. Bellaiche and Y. Ma,  Zeeman Effect in Centrosymmetric Antiferromagnetic Semiconductors
Controlled by an Electric Field, Phys. Rev. Lett. \textbf{129}, 187602 (2022).




\bibitem{p4}L. Webster and J. A. Yan, Strain-tunable magnetic anisotropy in monolayer $\mathrm{CrCl_3}$, $\mathrm{CrBr_3}$, and $\mathrm{CrI_3}$, Phys. Rev. B \textbf{98}, 144411 (2018).

\bibitem{p5}J. Liu, Q. Sun, Y. Kawazoe and P. Jena, Exfoliating biocompatible ferromagnetic Cr-trihalide monolayers,  Phys. Chem. Chem. Phys. \textbf{18}, 8777 (2016).

\bibitem{p6}M. Moaied, J. Lee and J. Hong, A 2D ferromagnetic semiconductor in monolayer Cr-trihalide and its Janus structures, Phys. Chem. Chem. Phys. \textbf{20}, 21755 (2018).
\bibitem{1}P. Hohenberg and W. Kohn, Inhomogeneous Electron Gas, Phys. Rev. \textbf{136},
B864 (1964); W. Kohn and L. J. Sham, Self-Consistent Equations Including Exchange and Correlation Effects, Phys. Rev. \textbf{140},
A1133 (1965).



\bibitem{pv1} G. Kresse, Ab initio molecular dynamics for liquid metals, J. Non-Cryst. Solids \textbf{193}, 222 (1995).

\bibitem{pv2} G. Kresse and J. Furthm$\ddot{u}$ller, Efficiency of ab-initio total energy calculations for metals and semiconductors using a plane-wave basis set, Comput. Mater. Sci. 6, \textbf{15} (1996).

\bibitem{pv3} G. Kresse and D. Joubert, From ultrasoft pseudopotentials to the projector augmented-wave method, Phys. Rev. B \textbf{59}, 1758 (1999).
\bibitem{pbe}J. P. Perdew, K. Burke and M. Ernzerhof, Generalized gradient approximation made simple, Phys. Rev. Lett. \textbf{77}, 3865 (1996).



\bibitem{bc} See Supplemental Material at [] for the related energy band structures;  the plane-averaged electrostatic potential.

\bibitem{mos} A. Y. Lu,  H. Y. Zhu,  J. Xiao et al., Janus Monolayers of Transition Metal Dichalcogenides. Nat. Nanotechnol. \textbf{12}, 744 (2017).

\bibitem{zg7}B. I. Weintrub, Y. L. Hsieh, S. Kovalchuk, J. N. Kirchhof, K. Greben, and K. I. Bolotin, Generating intense electric fields in 2D materials by dual ionic gating, Nat. Commun. 13, 6601 (2022).

\bibitem{ela}R. C. Andrew, R. E. Mapasha, A. M. Ukpong and N. Chetty, Phys. Rev. B \textbf{85}, 125428 (2012).

\bibitem{q5-1}K. N. Duerloo, M. T. Ong and E. J. Reed,  ntrinsic Piezoelectricity in Two-Dimensional Materials, J. Phys. Chem. Lett. \textbf{3}, 2871 (2012).


\bibitem{q5-1-1}C. Lee, X. g Wei, J. W. Kysar and J. Hone, Measurement of the elastic properties and intrinsic strength of monolayer graphene, Science \textbf{321}, 385 (2008).

\bibitem{gsd3}D. Lloyd, X. H. Liu, J. W. Christopher et al., Band Gap Engineering with Ultralarge Biaxial Strains in Suspended Monolayer $\mathrm{MoS_2}$,   Nano Lett. \textbf{16}, 5836 (2016).

   \bibitem{jpcm} H. Wang, V. Eyert and U. Schwingenschlogl, Electronic structure and magnetic ordering of the semiconducting chromium trihalides $\mathrm{CrCl_3}$, $\mathrm{CrBr_3}$, and $\mathrm{CrI_3}$, J. Phys.: Condens.
Matter \textbf{23}, 116003 (2011).


\end{references}
\end{document}